\def\lae{\mathrel{<\kern-1.0em\lower0.9ex\hbox{$\sim$}}}
\def\gae{\mathrel{>\kern-1.0em\lower0.9ex\hbox{$\sim$}}}

\newcommand{\be}{\begin{equation}}
\newcommand{\ee}{\end{equation}}

\documentclass[apj]{emulateapj}
\shorttitle{} \shortauthors{Fang \& Zhang}
\usepackage{apjfonts,lscape}

\begin{document}

\title{Multiband Nonthermal Radiative Properties of HESS J1813-178}
\author{Jun Fang \& Li Zhang}
\affil{Department of Physics, Yunnan University, Kunming, China}
\email{lizhang@ynu.edu.cn}
\begin{abstract}

The source HESS J1813-178 was detected in the survey of the inner
Galaxy in TeV $\gamma$-rays, and a composite supernova remnant (SNR)
G12.8-0.0 was identified in the radio band to be associated with it.
The pulsar wind nebula (PWN) embedded in the SNR is powered by an
energetic pulsar PSR J1813-1749, which was recently discovered.
Whether the TeV $\gamma$-rays originate from the SNR shell or the
PWN is uncertain now. We investigate theoretically the
multiwavelength nonthermal radiation from the composite SNR
G12.8-0.0. The emission from the particles accelerated in the SNR
shell is calculated based on a semianalytical method to the
nonlinear diffusive shock acceleration mechanism. In the model, the
magnetic field is self-generated via resonant streaming instability,
and the dynamical reaction of the field on the shock {\bf is} taken
into account. Based on a model which couples the dynamical and
radiative evolution of a PWN in a non-radiative SNR, the dynamics
and the multi-band emission of the PWN are investigated. The
particles are injected with a spectrum of a relativistic Maxwellian
plus a power law high-energy tail with an index of $-2.5$. Our
results indicate that the radio emission from the shell can be well
reproduced as synchrotron radiation of the electrons accelerated by
the SNR shock; with an ISM number density of $1.4$ cm$^{-3}$ for the
remnant, the $\gamma$-ray emission from the SNR shell is
insignificant, and the observed X-rays and very high energy (VHE)
$\gamma$-rays from the source are consistent with the emission
produced by electrons/positrons injected in the PWN via synchrotron
radiation and inverse Compton (IC) scattering, respectively; the
resulting $\gamma$-ray flux for the shell is comparable to the
detected one only with a relatively larger density of about $2.8$
cm$^{-3}$. The VHE $\gamma$-rays of HESS J1813-178 can be naturally
explained to mainly originate from the nebula although the
contribution of the SNR shell becomes significant with a denser
ambient medium.
\end{abstract}

\keywords{gamma rays: ISM--- ISM: individual objects (HESS
J1813-178, G12.8-0.0) - supernova remnants}

\section{Introduction}
\label{sec:intro}

The VHE source HESS J1813-178 was discovered with
the High Energy Stereoscopic System (HESS) in a survey of the inner
Galaxy in VHE $\gamma$-rays \citep[][]{A05}. The VHE $\gamma$-ray
image obtained with the HESS shows a pointlike source with an extension
of $\sim2.2'$, and the observed spectrum has a hard photon index
$2.09\pm0.08$ \citep[][]{A06}. HESS J1813-178 was also detected in
VHE $\gamma$-rays with the Major Atmospheric Gamma Imaging Cerenkov
(MAGIC) telescope. The differential flux given by MAGIC between 0.4 and
10 TeV can be well described by a power law with a index of
$-2.1\pm0.2$ \citep[][]{Al06}, which is consistent with the result obtained with
the HESS.

Firstly, HESS J1813-178 was unidentified, and it was assumed to be a
"dark particle accelerator" since no counterpart at lower
frequencies was reported ever. However, the shell of the SNR
G12.8-0.0 associated with HESS J1813-178 was discovered with a
diameter of $\sim2.5'$ in a new low-frequency Very Large Array (VLA)
90 cm survey \citep[][]{Bet05}. The non-thermal radio flux densities
of the shell-type SNR are $0.65\pm0.10$ and $1.2\pm0.08$ Jy at 20
and 90 cm, respectively \citep[][]{Bet05}. A highly absorbed X-ray source
AX J1813-178, for which the column density is $10^{23}$ cm$^{-2}$,
detected with {\it ASCA} is spatially coincident with the SNR. The
X-ray emission extending to 10 keV with a sharp cutoff below 2 keV
is primarily nonthermal, which can originate either from the SNR
shell or form the pulsar wind nebula (PWN) inside the remnant
\citep[][]{Bet05}. Due to the high column density derived from the
{\it ASCA} data, a distance of $\geq4$ kpc is derived for the source
\citep[][]{Bet05}.

Moreover, a soft $\gamma$-ray source, IGR J18135-1751, was
discovered as the counterpart of HESS J1813-178  \citep[][]{Uet05}.
It is persistent with a 20--100 keV luminosity of $5\times10^{34}$
erg s$^{-1}$ for a distance of 4 kpc. \citet[][]{Uet05} argued that
the observed properties of the source in the radio and X-ray bands
can be explained with the assumption that the source is a pulsar
wind nebula embedded in G12.8-0.0. Furthermore, high-angular
resolution X-ray observation with {\it XMM-Newton} shows that the
X-ray emitting object appears as a compact core located in the
center of the radio shell-type SNR G12.8-0.0 \citep[][]{Fet07}. They
argued that the source is a composite SNR since the central object
shows morphological and spectral resemblance to a PWN.

The observation with {\it Chandra} on the SNR G12.8-0.0 indicates
the X-ray source is a point surrounded by structured diffuse
emission that fills the interior of the radio shell
\citep[][]{Het07}. The compact source has a spectrum characterized
by a power law with an index of $\sim1.3$, typical of young and
energetic rotation-powered pulsars, and the morphology of the
diffuse emission strongly resembles that of a pulsar wind nebula
 \citep[][]{Het07}. Recently, an energetic pulsar PSR J1813-1749 with a period of
$\sim44.7$ ms, a characteristic age of 3.3 - 7.5 kyr, and a distance
of $4.7$ kpc by assuming the association with an adjacent young
stellar cluster, was discovered in a long, continuous {\it
XMM-Newton} X-ray timing observation \citep[][]{GH09}.  The pulsar
was found to be associated with the SNR G12.8-0.0, and it powers the
PWN \citep[][]{GH09}.

High-energy $\gamma$-rays can be produced either from SNR shells in
which particles are accelerated to relativistic through the first
Fermi process \citep[e.g.,][]{A05b,A07,BV06,FZ08,Fet09,MAB09}, or
from PWNe powered by the pulsars inside them \citep[e.g.,][]{Vet08,
ZCF08,GSZ09}. Although the source HESS J1813-178 is pointlike in VHE
$\gamma$-rays, the possibility of the VHE $\gamma$-rays originating
from the shell of the remnant cannot be ruled out given the size of
the SNR, the angular resolution of the HESS telescope, and the depth
of the observations \citep[][]{Al06}. In this paper, we study the
multiband nonthermal emission from the shell of the SNR G12.8-0.0
and the nebula inside it. The emission from the particles
accelerated by the SNR shock is investigated based on a
semianalytical method to the nonlinear diffusive shock acceleration
mechanism with a free escape boundary proposed by \citet{Cea10fre},
in which the amplified magnetic field due to resonant streaming
instability induced by cosmic rays and the dynamical feedback of
this self-generated magnetic field on the shock are taken into
account. On the other hand, the dynamics and the multi-band
radiative properties of the PWN are investigated basically according
to the model in \citet[][]{GSZ09}, which can self-consistently
describe the dynamical and radiative evolution of a pulsar wind
nebula in a non-radiative supernova remnant. Recently, based on the
long-term two dimensional particle-in-cell simulations,
\citet[][]{Sp08} found that the particle spectrum downstream of a
relativistic shock consists of two components: a relativistic
Maxwellian and a power law high-energy tail with an index of
$-2.4\pm0.1$. Different from \citet{GSZ09}, in which a single power
law injection spectrum for the electrons/positrons is employed to
discuss the radiative properties during different phase of the PWN,
in this paper we argue that the high-energy particles are injected
with the spectrum of a relativistic Maxwellian plus a power-law
high-energy tail during the evolution, and a kinetic equation is
used to obtain the energy distribution of the particles.

\section{Model and Results}
\label{sec:model}

In this section we describe the physics of our model for the
particles accelerated by a shock and its multiband nonthermal
emission (Section \ref{sec:shell}), the dynamics and multiwavelength
radiation of a PWN in the nonradiative shell (Section
\ref{sec:pwn}), and its implementation to the SNR G12.8-0.0 (Section
\ref{sec:app}).

\subsection{Particles accelerated by the SNR shock and its emission}
\label{sec:shell}

The pitch-angle averaged steady-state distribution of the protons
accelerated at a shock in one dimension satisfies the diffusive
transport equation \citep{MD01, B02, ABG08},
\begin{eqnarray}
\nonumber \frac{\partial}{\partial x}\left
[D\frac{\partial}{\partial x}f(x, p)\right ] &-& u\frac{\partial
f(x, p)}{\partial x} \\ &+& \frac{1}{3}\frac{du}{dx}p\frac{\partial
f(x, p)}{\partial p} + Q(x, p) = 0, \label{eq:diff}
\end{eqnarray}
where the coordinate $x$ is directed along the shock normal from
downstream towards upstream, $D$ is the diffusion coefficient and
$u$ is the fluid velocity in the shock frame, which equals $u_2$
downstream ($x>0$) and changes continuously upstream, from $u_1$
immediately upstream ($x=0^-$) of the subshock to $u_0$ at far
upstream. For the Bohm diffusion, $D=pc^2/(3eB)$, where $B$ is the
local magnetic field strength. The maximum momentum of the protons
accelerated by a SNR increases with time in the free-expansion phase
of the SNR due to the efficient magnetic field strength and the
constant shock speed. After the beginning of the Sedov-Taylor phase,
the maximum momentum drops with time due to the decrease of the
shock speed and the efficiency of the magnetic field amplification
\citep{Cea09,Cea10con}. In this case, particles with higher momentum
will escape from the SNR, and this phenomenon can be mimicked by
imposing a free escape boundary at a location upstream of the shock,
i.e., $f(-x_0, p)=0$ \citep{Cea10con,Cea10fre}.

With the assumption that the particles are injected at immediate
upstream of the subshock, the source function can be written as
$Q(x, p)=Q_0(p)\delta(x)$. For monoenergetic injection, $Q_0(p)$ is
\begin{equation}
Q_0(p) = \frac{\eta n_{\rm{gas, }1}u_1}{4\pi
p_{\rm{inj}}^2}\delta(p-p_{\rm{inj}})\;\; , \label{eq:Q0}
\end{equation}
where $p_{\rm{inj}}$ is the injection momentum, $n_{\rm{gas, }1}$ is
the gas density at $x=0^+$ and $\eta$ is the fraction of particles
injected in the acceleration process. With the injection recipe
known as thermal leakage, $\eta$ can be described as $\eta =
4(R_{\rm{sub}}-1)\xi^3e^{-\xi^2}/3\pi^{1/2}$ \citep{BGV05, ABG08},
where $R_{\rm{sub}}=u_1/u_2$ is the compression factor at the
subshock and $\xi$ is a parameter of the order of 2--4 describing
the injection momentum of the thermal particles in the downstream
region ($p_{\rm{inj}}=\xi p_{\rm{th,}2}$). $p_{\rm{th,}2}=
(2m_pk_{\rm{B}}T_2)^{1/2}$ is the thermal peak momentum of the
particles in the downstream fluid with temperature $T_2$, $m_p$ is
the proton mass and $k_{\rm{B}}$ is the Boltzmann constant. The
downstream temperature $T_2$ is calculated with equations (10) and
(11) in \citet{Cea09}. A relatively large $\xi=3.8 - 4.1$ is usually
employed to investigate the radiative properties of SNRs
\cite[e.g.,][]{MAB09,Mea09}, and we adopt $\xi=3.8$ in this paper.

The normalized pressure in cosmic rays is
\begin{equation}
P_c(x) = \frac{4\pi}{3\rho_0 u_0^2}\int_{p_{\rm inj}}^\infty dp p^3
v(p) f(x, p),   \label{eq:Pcx}
\end{equation}
where $\rho_0$ is the gas density far upstream of the shock.
Magnetic fields can be generated by streaming instability induced by
the accelerated particles. With the assumption of the turbulence is
generated by the resonant streaming instability, the normalized
pressure of the amplified magnetic field can be described as
\citep{Cea09,Cea10fre}
\begin{equation}
P_w(x) = \frac{v_{A}}{4u_0}\frac{1 - U^2(x)}{U^{3/2}(x)},
\label{eq:Pwx}
\end{equation}
where $U(x) = u(x)/u_0$, $v_{A}=B_0/(4\pi\rho_0)^{1/2}$ is the
Alfv\'{e}n velocity, $B_0$ is the background magnetic field
strength. Then the strength of the amplified field is $\delta B(x) =
(8\pi \rho_0 u_0^2P_w(x))^{1/2}$, and the magnetic field downstream
of the shock is further enhanced by $B_2 = R_{\rm sub} B_1$
\citep{MAB09}, where $B_1 = \delta B(0)$ is the amplified magnetic
field immediately upstream of the subshock. The effect of turbulent
heating is ignored in this paper because the properties of the
damping of the magnetic field is still uncertain now, and a new
parameter is needed even with the damping phenomenologically taken
into account \citep{Cea10con}. Neglecting the effect of turbulent
heating, the normalized pressure of the background gas is
\begin{equation}
P_g(x) = \frac{U(x)^{-\gamma}}{\gamma M_0^2}, \label{eq:Pgx}
\end{equation}
where $M_0$ is the sonic Mach number of the shock, $\gamma=5/3$ is
the adiabatic index. The total compression factor $R_{\rm tot}$ is
related with the compression factor at the subshock $R_{\rm sub}$
through \citep{Cea09}
\begin{equation}
R_{\rm tot}^{\gamma + 1} = \frac{M_0^2R_{\rm sub}^{\gamma}}{2}
\left[\frac{\gamma + 1 - R_{\rm sub}(\gamma - 1)}{1 + \Lambda_B}
\right],
\end{equation}
where
\begin{equation}
\Lambda_B = \frac{P_w(0)}{P_g(0)}[1 + R_{\rm sub}(2/\gamma - 1 )].
\end{equation}

\begin{figure}
\includegraphics[scale=0.7]{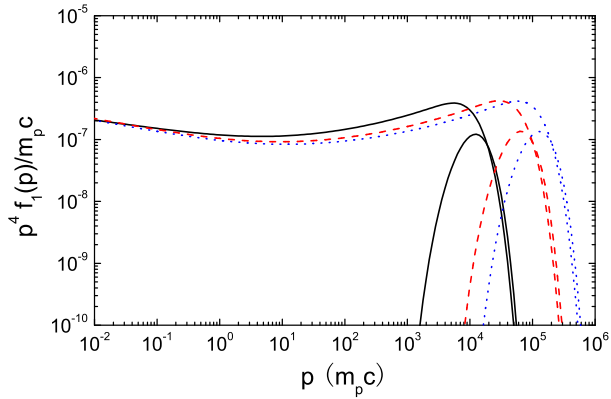}
\caption{\label{fig1} Particle spectra at the shock and the escape
fluxes ($p^4 \phi_{\rm esc}/u_0$) with $\xi = 3.8$, $T_0=10^4$ K,
$n_{0}=1.4$ cm$^{-3}$, $B_0=5\,\mu$G, $u_0=1\times10^8$ cm s$^{-1}$,
$x_0 = 0.18$ pc (solid line); $x_0 = 0.9$ pc (dashed line); and $x_0
= 1.8$ pc (dotted line).}\label{fig:new1}
\end{figure}

\begin{figure}
\includegraphics[scale=0.7]{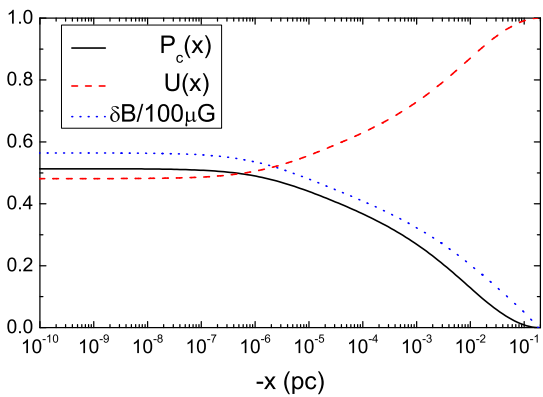}
\caption{\label{fig1}  $P_c(x)$ (solid line), $U(x)$ (dashed line),
and $\delta B/100\,\mu$G (dotted line) in the upstream region for
$x_0 = 0.9$ pc. Others are the same as Fig.\ref{fig:new1}.
}\label{fig:new2}
\end{figure}

Given a value of $\xi$, a temperature far upstream of the shock
$T_0$, a shock velocity $u_0$, a background magnetic field $B_0$,
and $x_0$, the particle spectrum at the shock $f_1(p)$ and the
escape flux $\phi_{\rm esc}(p) = -[D(x,p)\partial f/\partial
x]_{x_0}$ can be obtained with the method proposed in
\citet{Cea10fre}, in which the diffusion coefficient is calculated
in the self-generated magnetic field induced by resonant streaming
instability of the accelerated particles. In order to show how the
protons spectrum is affected by the choice of $x_0$,  we plot in
Fig.\ref{fig:new1} the particle spectra at the sub-shock position
and the escaping fluxes for three different values of $x_0$: 0.18 pc
(solid line); 0.9 pc (dashed line); and = 1.8 pc (dotted line). The
other parameters are all fixed to: $\xi = 3.8$, $T_0=10^4$ K,
$n_{0}=1.4$ cm$^{-3}$, $B_0=5\,\mu$G, $u_0=1\times10^8$ cm s$^{-1}$.
The particle spectrum for each $x_0$ is cut off at a maximum
momentum $p_{\rm max}$, which increases with the value of $x_0$.
$U(x)$, $P_c(x)$ and $\delta B$ in the upstream region for $x_0 =
0.9$ pc are indicated in Fig.\ref{fig:new2}. The amplified magnetic
field immediately upstream of the subshock $B_1$ is 56 $\mu G$,
which is significantly stronger than the background magnetic field.

The electrons have the same spectrum of the protons up to a maximum
energy determined by synchrotron losses. Note that the spectrum of
the  accelerated electrons at the shock around the cutoff momentum
$p_{{\rm max}, e}$ is difficult to obtain in a fully nonlinear
scenario. With the test particle approximation, the energy spectrum
of electrons accelerated by SNR shocks for a strong shock can be
described as \citep[][]{ZA07,B10}
\begin{equation}
\label{pemaxza}
f_e(x, p)=K_{ep}f(x, p)\left[1+0.523(p/p_{{\rm max},
e})^2\right]\exp(-p^2/p_{{\rm max}, e}^2). \label{fenew}
\end{equation}
However, the spectrum cut off by a simple exponential is also
employed to reproduce the multiband emission from SNRs \citep[e.g.,][]{Fet09}, and
we use this form in this paper, i.e.,
\begin{equation}
f_e(x, p)=K_{ep}f(x, p)\exp(-E(p)/E_{{\rm max}, e}), \label{fe}
\end{equation}
where $E(p)$ is the kinetic energy of the electrons, and the
electron/proton ratio $K_{ep}$ is treated as a parameter. The choice
of the cut-off shape of the spectrum has no influence on our results
because in this paper we interpret the observed X-ray from emission
G12.8-0.0 as due to the PWN inside the remnant. The maximum energy
of electrons results equating the synchrotron loss time with the
acceleration time. In the context of non linear shock acceleration
an approximate solution is given by the following equation
\citep[][]{Bet02}
\begin{equation}
E_{{\rm max}, e} = 6\times10^7 \left(\frac{u_0}{10^8 {\rm cm\
s^{-1}}}\right)\left[\frac{R_{\rm tot}-1}{R_{\rm tot}(1+R_{\rm
sub}R_{\rm tot})}\left(\frac{10\mu {\rm G}}{B_1}\right)\right]^{1/2}
{\rm MeV} ,
\end{equation}
where the magnetic field immediately upstream of the shock, $B_1$,
is assumed to be compressed by a factor $R_{\rm sub}$.

Assuming the accelerated particles distribute homogeneously and most
of the emission is from downstream of the shock, and using the
distribution function at the shock to represent the particle
distribution in the whole emitting zone, the volume-averaged
emissivity for photons produced via p-p interactions can be written
as
\begin{equation}
Q(E)=4\pi n_{\rm gas}\int dE_{\rm p} J_{\rm p}(E_{\rm
p})\frac{d\sigma(E, E_{\rm p})}{dE} \;\; , \label{eq:PP}
\end{equation}
where $E_{\rm p}$ is the proton kinetic energy, $n_{\rm gas}=R_{\rm
tot}n_{\rm 0}$ is the gas number density downstream of the shock,
and $J_{\rm p}(E_{\rm p})=v p^2 f_0(p) dp/dE_{\rm p}$ is the
volume-averaged proton density and $v$ is the particles' velocity.
We use the differential cross-section for photons $d\sigma(E, E_{\rm
p})/dE$ presented in \citet{Ka06} to calculate the hadronic
$\gamma$-rays produced via p-p interaction. Finally, the photon flux
observed at the earth can be obtained with
\begin{equation}
F(E)=\frac{VQ(E)}{4\pi d^2} \;\; , \label{eq:Flux}
\end{equation}
where $d$ is the distance from the earth to the source and $V$ is
the average emitting volume of the source. For the accelerated
protons, the emitting volume can be estimated by
$V_p\approx(4\pi/3)R_{\rm snr}^3 /R_{\rm tot}$ \citep{E00}, here
$R_{\rm snr}$ is the radius of the SNR.  For the electrons, the
thickness of the emitting region can be estimated by solving the
diffusion-convection equation downstream of the shock, i.e., $u_2
(\partial f_e/\partial x)=D\partial^2/\partial x^2 - f_e/\tau_{\rm
syn}$, where $\tau_{\rm syn}$ is the time scale of the synchrotron
radiation \citep[][]{MAB09}. The solution shows $f_{e}\propto
\exp(-x/R_{\rm rim})$, and the spatial scale $R_{\rm rim}$ is given
by \citep[][]{BV04}
\begin{equation}
R_{\rm rim}(p) = \frac{2D_2/u_2}{\sqrt{1+4D_2/u_2\tau_{\rm syn}}-1
}.
\end{equation}
As a result of the synchrotron losses, electrons with relatively
high energy are confined into a thin rim behind the shock and the
total emitting volume is smaller than $V_{\rm p}$. A break in the
spectrum occurs for $p = p_{\rm t}$, defined as the momentum where
the time scale of the losses equals to the age of the remnant, i.e.
$\tau_{\rm syn}(p_{\rm t})= t_{\rm snr}$. In the steady state, the
volume of the electrons with higher energies is smaller than that of
the protons due to the strong synchrotron losses. Hence $V_e(p) =
V_p$ for $p\leq p_{\rm t}$, while $V_e(p)=V_{\rm p}R_{\rm
rim}(p)/(R_{\rm snr}/3R_{\rm tot})$ for $p> p_{\rm t}$, where
$p_{\rm t}$ is determined by $4\pi R_{\rm snr}^2 R_{\rm rim}(p) =
V_p$.

\subsection{Dynamics and radiative properties of a PWN inside a
nonradiative SNR } \label{sec:pwn}

A PWN is powered by the pulsar which dissipates its rotational
energy into the nebula. The spin-down luminosity of a pulsar with a
rotation period of $P$ evolves with time as
\citep[e.g.,][]{GS06,S08}
\begin{equation}
\dot{E}(t)=\dot{E}_0\left(1+\frac{t}{\tau_0}
\right)^{-\frac{n+1}{n-1}}, \label{Lum}
\end{equation}
where $\tau_0$ is the spin-down time scale of the star, $\dot{E}_0$
is the initial spin-down power, $n$ is the braking index of the
pulsar, which is equal to 3 for magnetic dipole spin-down.

High-energy particles and magnetic fluxes are injected into the PWN
from the terminate shock located where the ram pressure of the
unshocked wind is equal to that of the nebula. In this paper, we
assume the spin-down power is distributed between electrons and
positrons ($\dot{E}_{\rm e}(t) = \eta_{\rm e}\dot{E}(t)$), and
magnetic fields ($\dot{E}_{\rm B}= \eta_{\rm B}\dot{E}(t))$
\citep[e.g.,][]{GSZ09}. In \citet[][]{GSZ09}, they used a simple
power-law injection spectrum for the electrons/positrons to discuss
the radiative properties during different phase of the PWN. However,
a broken power-law spectrum is usually needed to reproduce the
non-thermal emission of a PWN with multi-band observations
\citep[e.g.,][]{AA96,Vd06,Set08,ZCF08}. Recently, based on the
long-term two-dimensional particle-in-cell simulations,
\citet[][]{Sp08} found that the particle spectrum downstream of a
relativistic shock is a Maxwellian plus a power-law tail with a
index of $2.4\pm0.1$, and this spectrum also was used to investigate
the multiband emission from PWNe \citep[e.g.,][]{FZ10aa,Slea10}. We
assume high-energy particles injected in a PWN are accelerated by
the termination shock (TS) which is typically relativistic with a
lorentz factor $\sim10^6$ with respect to the pulsar wind upstream
of the shock. Therefore, we assume the spectrum of the high-energy
particles injected in the PWN has the form,
\begin{equation}
Q(E,t) = \left\{\begin{array}{ll} C(t)\frac{E}{E_{\rm b}} \exp\left(-\frac{E}{E_{\rm b}}\right) & E\leq E_{\rm min}\\
C(t)\left[\frac{E}{E_{\rm b}}\exp\left(-\frac{E}{E_{\rm
b}}\right)+f\left(\frac{E}{E_{\rm b}}\right)^{-\alpha}\right] &
E_{\rm min}<E \leq E_{\rm max}
\end{array}
\right . , \label{dnde}
\end{equation}
where, $\alpha=2.4\pm0.1$, $E_{\rm b}\sim2.6\times10^5 \gamma_{\rm
ts, 6}$ MeV, $\gamma_{\rm ts, 6}$ is the Lorentz factor of the
upstream pulsar wind of the TS in units of $10^6$, $E_{\rm min} =
f_{\rm min}E_{\rm b}$ with $f_{\rm min}\sim7$, $f$ is normalized by
$E_{\rm min}/E_{\rm b}\exp\left(-E_{\rm min}/E_{\rm
b}\right)=f\left(E_{\rm min}E_{\rm b}\right)^{-\alpha}$. $C(t)$ can
be obtained with
\begin{equation}
C(t) = \frac{\dot{E}_{\rm e}(t)}{2E_{\rm b}^2 + f\frac{E_{\rm
b}^2}{2-\alpha} \left[ \left( \frac{E_{\rm max}}{E_{\rm b}}
\right)^{2-\alpha} - \left( \frac{E_{\rm min}}{E_{\rm b}}
\right)^{2-\alpha}\right]}. \label{C}
\end{equation}
Assuming the particles are homogeneously distributed in the PWN, and
the energy distribution of these particles in the nebula evolves as
\begin{equation}
\frac{\partial N(E,t)}{\partial t}= \frac{\partial }{\partial E}
\left [ \dot{E} N(E,t) \right ] + Q(E,t) , \label{Distr}
\end{equation}
where $\dot{E}$ is the energy-loss rate of the particles with energy
$E$. Energy-loss mechanisms include synchrotron radiation, IC
scattering and adiabatic loss.

\begin{figure}
\includegraphics[scale=1.2]{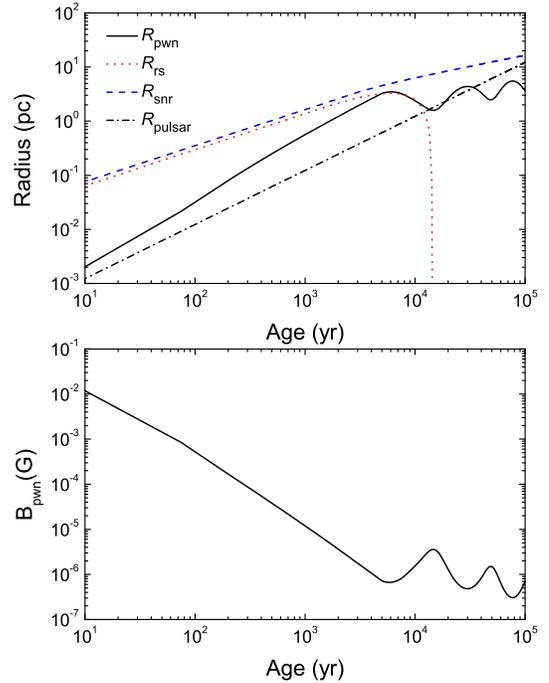}
\caption{\label{dynfig} Upper panel: radius of the SNR ($R_{\rm snr}$, dashed
line), reverse shock ($R_{\rm rs}$, dotted line), PWN ($R_{\rm
pwn}$, solid line), and the location of the pulsar ($R_{\rm
pulsar}$, dash-dotted line). Lower panel: magnetic field
strength in the PWN ($B_{\rm pwn}$).}
\end{figure}

The dynamics of the PWN inside the supernova shell is calculated
basically following the model presented in \citet[][]{GSZ09}. The
model assumes that the progenitor supernova ejects material with
mass $M_{\rm ej}$ and energy $E_{\rm sn}$ into the ambient matter
with a constant density $\rho_{\rm ISM}$. Assuming the PWN has no
influence on the dynamics of the forward shock and the reverse
shock, the velocity and the radius of the forward shock and the
reverse shock of the surrounding SNR are calculated with the
equations in \citet[][]{TM99}. The pulsar dissipates energy into the
PWN, which sweeps up the supernova ejecta into a thin shell
surrounding the nebula, and new particles are injected into the PWN
at each time step.

The dynamical evolution of the PWN with the parameters for the SNR
G12.8-0.0 (see Table\ref{para}) is shown in Fig.\ref{dynfig}. The
radiuses of the SNR ($R_{\rm snr}$), the reverse shock ($R_{\rm
rs}$), the PWN ($R_{\rm pwn}$) and the pulsar are indicated by the
dashed, dotted, solid and dash-dotted lines, respectively. A
velocity of 120 km s$^{-1}$ is used in the calculation, which has no
influence on the resulting dynamical structure and radiative
property of G12.8-0.0 because the young pulsar is safely in the
nebula now (see Section \ref{sec:app}). Initially, the pressure of
the PWN is much bigger than the pressure of the surrounding
supernova ejecta, so the PWN expands adiabatically into the cold
supernova ejecta. The ejecta surrounding the PWN is swept up to a
thin shell, which is decelerated by ram pressure since its velocity
is bigger than the local sound speed \citep[][]{Get07}. The mass of
the PWN $M_{\rm pwn}$ increases continuously since the shell expands
faster than the ambient ejecta. This expansion phase ends when the
PWN collides with the reverse shock of the SNR. After the collision,
the pressure inside the nebula $P_{\rm pwn}$ is much smaller than
the pressure of the material around it $P_{\rm snr}(R_{\rm pwn})$.
The velocity of the PWN decreases greatly, and finally the PWN is
compressed. During this process of compression, the magnetic field
strength in the nebula increases significantly. Furthermore, the
radius of the PWN decrease significantly, and the PWN will expand
again when the inner pressure eventually becomes bigger than that of
the surrounding ejecta. The nebula experiences a series of
contractions and re-expansions until the SNR enters the radiative
phase of its evolution. The pulsar moving in the space will leave
the PWN during the compression, and it can re-enter the nebula when
the nebula expands again. With the parameters in Table.\ref{para},
the pulsar firstly leaves the nebula at $\sim13700$ yr, and
re-enters it at $\sim19000$ yr.

\subsection{Application to G12.8-0.0 and discussion}
\label{sec:app}

\begin{table}
\begin{center}
\caption{Input Parameters for the SNR G12.8-0.0 \label{para}}
\begin{tabular}{lccc}
\hline \hline
{\sc Parameter} & {\sc Value} & {\sc Parameter} & {\sc Value} \\
\hline $d$      & 4.7 kpc         & $\dot{E_0}$ & $4.2\times10^{38}$ erg s$^{-1}$ \\
$E_{\rm sn}$    & $0.4\times10^{50}$erg    & $\tau_{\rm sd}$ & 500 yr            \\
$M_{\rm ej}$    & 3$M_{\odot}$    &  $n$             & 3.0                \\
$n_{\rm 0}$   & 1.4 cm$^{-3}$     & $\eta_{\rm B}$  & $1\times10^{-4}$             \\
Age             & 1200 yr         &  $E_{\rm max}$   & 1000 TeV    \\
$T_{0}$ & $10^4$ K           & $E_{\rm b}$     & $3\times10^5$ MeV    \\
$B_{0}$   & 5 $\mu$G       & $v_{\rm psr}$   & 120  km s$^{-1}$            \\
$K_{ep}$        & $1.8\times10^{-3}$            &    &          \\

\hline
\end{tabular}
\end{center}
\end{table}

\begin{figure}
\includegraphics[scale=1.2]{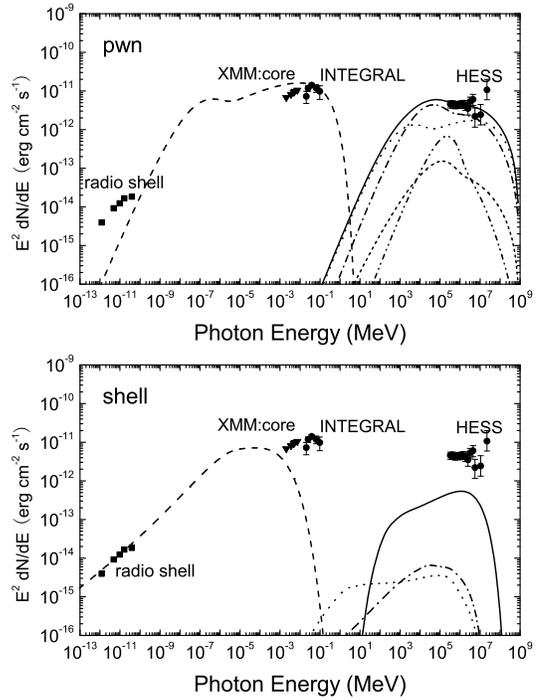}
\caption{\label{g128} Upper panel: the spectral energy distribution
for the PWN in G12.8-0.0 at an age of 1200 yr. Synchrotron radiation
(dashed line), and IC scattering on the CMB (dotted line), IR
(dash-dotted line), star light (dash-dot-dotted line) and the
synchrotron photons (short dashed line) are shown in the figure. The
solid line represents the whole IC scattering on all the soft
photons. Lower panel: the spectral energy distribution of the
emission from the shell of the SNR G12.8-0.0. Synchrotron radiation
(dashed line), bremsstrahlung (dotted line) IC scattering
(dash-dotted line) of the accelerated electrons and p-p interactions
(solid line) of the accelerated protons are demonstrated in the
figure. The radio data are from the VLA, Bonn, Parkes, and Nobeyama
observatories \citep[][]{Bet05}. The X-ray data are from {\it
XMM-Newton} \citep[][]{Fet07} and {\it INTEGRAL} \citep[][]{Uet05}.
The HESS flux points in the VHE $\gamma$-ray band are from
\citet[][]{A06}. }
\end{figure}

\begin{figure}
\includegraphics[scale=0.7]{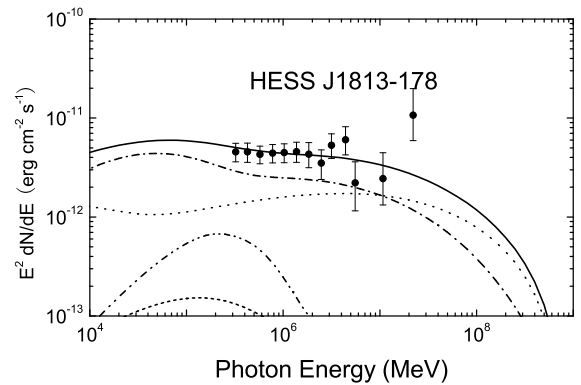}
\caption{\label{g128clear} A zoomed view of the resulting spectrum
and the HESS results in the TeV band. Others are the same as the
upper panel in Fig.\ref{g128}.}
\end{figure}

The shell with a diameter of $\sim2.5'$, corresponding to a radius
of 1.7 pc for a distance of 4.7 kpc, of the SNR G12.8-0.0 had been
revealed with the radio observations with a spectral index of
$\sim-0.48$ \citep[][]{Bet05}. Moreover, a pulsar wind nebula with a
diameter of $\sim80''$ embedded in the SNR was been disclosed in the
observations from the X-ray to soft $\gamma$-ray bands
\citep[][]{Uet05,Fet07,Het07}. An energetic pulsar with a spin-down
power bigger than $10^{37}$ erg s$^{-1}$ is argued to exist in the
nebula based on the ASCA and Chandra observations \citep[][]{Het07}.
Recently, the energetic pulsar PSR J1813-1749 associated the PWN was
discovered in the long, continuous {\it XMM-Newton} X-ray timing
observation, and the pulsar has a period of $\sim44.7$ ms, a
spin-down age of 3.3 -- 7.5 kyr, and a spin power of $6.8\pm2.7
\times10^{37}$ erg s$^{-1}$ \citep[][]{GH09}.

The small radius places the SNR G12.8-0.0 amongst the smallest known
SNRs, which also suggests a young age for the remnant. In the model,
the radius of the SNR is determined by the explosion energy, the
mass of the ejecta, the ambient density, and the age of the system.
We find that, with $E_{\rm sn}=0.4\times10^{50}$ erg, $M_{\rm
ej}=3M_{\odot}$, and an ambient ISM density $n_{0}=1.4$ cm$^{-3}$,
the resulting radius of the SNR shell is $\sim1.8$ pc at an age of
$\sim 1200$ yr, which is consistent with the radius of the radio
shell. With a larger explosion energy ($E_{\rm sn}\geq10^{50}$ erg)
or a smaller ambient density ($n_{0}\leq 0.1$ cm$^{-3}$), the age of
the system, which is constrained by the radius of the SNR, is much
smaller the radius of the radio shell, and the resulting flux of
X-ray emission from the PWN is usually much higher than the observed
one by {\it XMM-Newton} \citep[][]{Fet07} and {\it INTEGRAL}
\citep[][]{Bet05} since the magnetic field is stronger for a smaller
age (see the lower panel of Fig.\ref{dynfig}). A sub-energetic
explosion with $E_{\rm sn}\leq10^{50}$ erg and a low ejecta mass
$M_{\rm ej}=3M_{\odot}$ was also argued to exist from the
multiwavelength study of the newly discovered SNR G310.0-1.6
\citep{Rea09}. With $\tau_{\rm sd}=500$ yr and
$\dot{E_0}=4.2\times10^{38}$ erg s$^{-1}$, the spin-down power is
$3.6\times10^{37}$ erg s$^{-1}$ at an age of 1200 yr from
Eq.\ref{Lum}, which is a little smaller than the present observed
value, i.e., $6.8\pm2.7\times10^{37}$ erg s$^{-1}$ \citep[][]{GH09}.
The radius of the PWN at this age is 0.7 pc, and the ratio $R_{\rm
pwn}/R_{\rm snr}$ is $\sim0.4$. The resulting multiband radiation
from the PWN in the composite SNR G12.8-0.0 at an age of 1200 yr is
shown in the upper panel of Fig.\ref{g128} with the parameters in
Table\ref{para}. In the figure we also plot the observed data from
radio to TeV band for comparison. The data sources are reported in
the figure caption. For each panel, the interstellar soft photons
scattered in the IC process contain the CMB, IR ($T_{\rm IR}=35$ K,
$U_{\rm IR}=1.0$ eV cm$^{-3}$) and star light ($T_{\rm st}=3000$ K,
$U_{\rm st}=1.5$ eV cm$^{-3}$). The observed emission detected with
{\it XMM-Newton} and {\it INTEGRAL} from the X-ray to hard X-ray
bands can be explained as the synchrotron radiation from the
electrons injected in the PWN, although the resulting flux is a
little higher than that observed with {\it XMM-Newton}. In the VHE
$\gamma$-ray band, the photons are mainly produced via IC scattering
of the electrons in the nebula on the CMB and IR photons from
interstellar dust (Fig.\ref{g128clear}).

\begin{figure}
\includegraphics[scale=0.7]{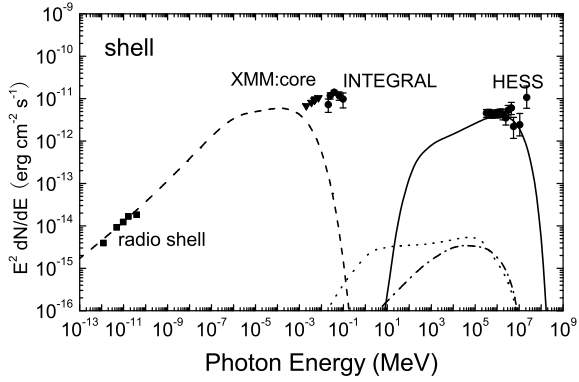}
\caption{\label{figs6}  the spectral energy distribution of the
emission from the shell for $n_0=2.8$\,cm$^{-3}$ and
$K_{ep}=4.5\times10^{-4}$. The others are the same as the lower
panel of Fig.\ref{g128}. }
\end{figure}

The upstream temperature for a SNR varies between $\sim 10^4$ K for
a typical ISM up to $\sim10^7$ K if the SNR expands in the hot
bubble generated by the progenitor's wind
\citep[][]{CL89,MAB09,BV10}. We assume the SNR G12.8-0.0 is
expanding in the typical ISM with a temperature of $\sim 10^4$ K to
investigate the multiband nonthermal emission from the SNR shell. At
an age of 1200 yr, the velocity of the SNR shock is $\sim1000$ km
s$^{-1}$, corresponding to a Mach number of $\sim 86$ for $T_0 =
10^4$ K. With these parameters, $R_{\rm tot}$ and $R_{\rm sub}$ are
about 8.1 and 3.55, respectively, and the downstream temperature
$T_2$ is $3.8\times10^6$ K. Moreover, the maximum energies of the
accelerated protons and electrons are about 60 TeV and 4 TeV,
respectively. The energy contained in the protons accelerated by the
shock is $5\times10^{48}$ erg, and it in the electrons is
$6\times10^{45}$ erg. Therefore, about $12\%$ of the explosion
energy $E_{\rm sn}$ has been converted to the kinetic energy of the
particles. The amplified magnetic field strength immediately
upstream of the subshock is 62\,$\mu$G, and then the downstream
magnetic field strength is 220\,$\mu$G. With this downstream
magnetic field strength and $K_{ep}=1.8\times10^{-3}$, the flux
points of the radio shell can be reproduced as synchrotron radiation
of the electrons accelerated by the SNR forward shock wave; whereas
the flux from p-p collisions of the accelerated protons on the
ambient matter is significantly smaller than that observed with
HESS.

The ISM number density $n_0$ is an important parameter which can
greatly influence the multiband nonthermal emission from the SNR
shell. With a larger $n_0$, both the density of the accelerated
particles and that downstream of the shock increase accordingly. As
a result, the flux of the p-p collisions is enhanced with $\propto
n_0^2$, and then a smaller $K_{ep}$ is usually needed to reproduce
the observed radio fluxes for a denser medium. In order to reproduce
the HESS flux in the TeV band via p-p collisions from protons
accelerated by the SNR shell, a denser medium with a density of
$2.8$ cm$^{-3}$ must be used in the model (see Fig.\ref{figs6}). For
$n_0 = 2.8$ cm$^{-3}$ and the other parameters the same as
Fig.\ref{g128}, the multiband nonthermal emission from the PWN is
nearly the same as that for $n_0 = 1.4$ cm$^{-3}$, whereas the
$\gamma$-ray flux for the p-p collisions in the SNR shell is
comparable to the HESS result; the downstream magnetic field
strength is 271\,$\mu$G, and then a smaller
$K_{ep}=4.5\times10^{-4}$ is needed to reproduced the observed
fluxes in the radio band. In this SNR-dominated scenario, the
$\gamma$-ray photon index of the resulting emission is $<2$ up to
several tens of GeV, whereas it is $>2$ at higher energies
(Fig.\ref{figs6}). On the other hand, the index in the PWN-dominated
case is $<2$ from 0.1 GeV to 1 TeV (Fig.\ref{g128}). The resulting
TeV $\gamma$-ray spectral energy distributions in the two scenarios
are all consistent with the HESS flux points, so we cannot determine
which case is preferred by comparison with the detected fluxes in
the TeV range now. However, the flux in the GeV band in the
PWN-dominated case is several times higher than in the SNR-dominated
case, so future detections in this band can give constraints on the
origin of the $\gamma$-rays from HESS J1813-178.

Even if a denser medium with $n_0 \sim 2.8$ cm$^{-3}$ is used in the
model to enhance the p-p collisions in the SNR shell, the IC
scattering of the electrons/positrons in the PWN is still prominent
enough to contribute significantly to the high-energy $\gamma$-ray
emission from the composite SNR. In the model, the PWN evolves in
the remnant with an age of $\sim 1200$ yr, and high-energy
electrons/positrons are injected continuously into the nebula from
the energetic pulsar during the evolution. In fact, if the emission
detected with {\it XMM-Newton} and {\it INTEGRAL} is the synchrotron
radiation from the electrons/positrons injected in the PWN, the
detected VHE $\gamma$-rays can be easily explained by the IC
scattering of these electrons/positrons off the ambient soft
photons. Therefore, it is natural to argue that the observed
high-energy photons from the X-ray to VHE $\gamma$-ray bands
originate mainly from the PWN inside the composite remnant although
the contribution of the SNR shell to the $\gamma$-rays becomes
significant for a denser medium with a density of $2.8$ cm$^{-3}$.

\section{Summary and conclusion}
\label{sec:discussion}

The origin of the VHE $\gamma$-rays from HESS J1813-178 is
investigated in this paper. Although the source is pointlike in VHE
$\gamma$-rays, a shell origin of the  VHE $\gamma$-rays cannot be
ruled out because the PSF of the HESS telescope is very close to the
size of the SNR's shell \citep[e.g.,][]{Al06}. A SNR with a diameter
of $2.5'$ was identified in the radio bands to be associated with
the VHE $\gamma$-ray source. Moreover, X-ray observations showed a
PWN powered by an energetic pulsar is embedded in the SNR
\citep[][]{Bet05,Uet05,Fet07,Het07}, and the pulsar was recently
discovered in a long, continuous {\it XMM-Newton} X-ray timing
observation \citep[][]{GH09}. We apply a model which can
self-consistently calculate the multiwavelength nonthermal emission
both from the SNR shell and from the PWN embedded in the remnant. In
the model, electrons/positrons with a spectrum of a relativistic
Maxwellian plus a power-law high-energy tail are injected in the
nebula during the PWN during its evolution inside the SNR; protons
and electrons are accelerated by the SNR shock wave, and the
spectrum of the accelerated particles are calculated with a
semi-analytical non-linear model. Our results indicate that: with
the parameters in Table\ref{para}, (1) the observed emission of the
shell in the radio bands can be well explained as synchrotron
radiation of the electrons accelerated by the SNR shock wave,
whereas the flux of p-p collisions of the accelerated protons is
significantly smaller than the TeV flux observed with HESS; (2) the
observed emission in the X-ray to hard X-ray band detected with {\it
XMM-Newton} and {\it INTEGRAL}, respectively, can be explained as
the synchrotron radiation of the electrons/positrons injected in the
PWN; (3) the VHE $\gamma$-rays in the TeV band are mainly produced
via IC scattering of the electrons/positrons injected in the nebula
on the CMB and interstellar IR photons.

With a sub-energetic explosion ($E_{\rm sn}\sim 0.4\times 10^{50}$
erg), an age of $\sim 1200$ yr, and an ambient density of $1.4$
cm$^{-3}$, the radius of the composite SNR G12.8-0.0, 1.7 pc, as
well as the multiband observed nonthermal fluxes for the remnant can
be reproduced within this model described in this paper. The present
value of the SNR shock's velocity is about 1000 km s$^{-1}$, and the
VHE $\gamma$-rays are predominately produced from the nebula inside
the remnant via IC scattering. Of course, p-p interaction from the
SNR shell can be enhanced with a denser medium around the remnant
(Fig.\ref{figs6}). Even if this condition were satisfied, the
$\gamma$-rays produced via IC scattering in the PWN remain
significant, hence, in this case, both the SNR shell and the PWN
would contribute to the production of the observed $\gamma$-ray
flux. In GeV $\gamma$-rays, the spectral property of the resulting
emission in this SNR-dominated scenario (Fig.\ref{figs6}) differs
significantly from that in the PWN-dominated case (Fig.\ref{g128}).
The VHE $\gamma$-rays of HESS J1813-178 detected with HESS can be
naturally explained as the IC scattering of the electrons/positrons
injected into the PWN although the p-p collisions become important
with a denser ambient medium for the remnant. Our study give more
insights on the nature of the multiband nonthermal emission of the
composite SNR G12.8-0.0, even though some assumptions are made in
the model.

\section*{Acknowledgments}
We are very grateful to the anonymous referee for his/her helpful
comments to improve the paper. This work is partially supported by
the Scientific Research Foundation of Graduate School of Yunnan
University,  the National Natural Science Foundation of China (NSFC
10778702, 10803005), a 973 Program (2009CB824800), and Yunnan
Province under a grant 2009 OC.



\begin{thebibliography}{00}


\bibitem[Aharonian et al.(2005a)]{A05}
Aharonian, F., et al. (HESS Collaboration) 2005a, Science, 307, 1938

\bibitem[Aharonian et al.(2005b)]{A05b}
Aharonian, F., et al. (HESS Collaboration) 2005b, A\&A, 437, L7

\bibitem[Aharonian et al.(2006)]{A06}
Aharonian, F., et al. (HESS Collaboration) 2006, \apj, 636, 777

\bibitem[Aharonian et al.(2007)]{A07}
Aharonian, F. et al. (HESS Collaboration), 2007, A\&A, 464, 235

\bibitem[Albert et al.(2006)]{Al06}
Albert, J. et al. 2006, \apj, 637, L41


\bibitem[Amato, Blasi \& Gabici(2008)]{ABG08}
Amato, E., Blasi, P., \& Gabici, S., 2008, MNRAS, 385, 1946

\bibitem[Atoyan \& Aharonian(1996)]{AA96}
Atoyan, A. M., \& Aharonian, F. A. 1996, MNRAS, 278, 525

\bibitem[Berezhko et al.(2002)]{Bet02}
Berezhko, E. G., Ksenofontov, L. T. \&  V\"{o}lk, H. J. 2002, A\&A,
395, 943

\bibitem[Berezhko \& V\"{o}lk(2004)]{BV04}
Berezhko, E. G., \& V\"{o}lk, H. J., 2006, A\&A, 419, L27

\bibitem[Berezhko \& V\"{o}lk(2006)]{BV06}
Berezhko, E.G. \& V\"{o}lk, H.J. 2006, A\&A, 451,981


\bibitem[Berezhko \& V\"{o}lk(2010)]{BV10}
Berezhko, E.G. \& V\"{o}lk, H.J. 2010, A\&A, 511. 34

\bibitem[Blasi(2002)]{B02}
Blasi P., 2002, Astropart. Phys., 16, 429

\bibitem[Blasi(2010)]{B10}
Blasi, P., 2010, MNRAS, 402, 2807

\bibitem[Blasi, Gabici \& Vannoni(2005)]{BGV05}
Blasi P., Gabici S., \& Vannoni G., 2005, \mnras, 361, 907

\bibitem[Brogan et al.(2005)]{Bet05}
Brogan, C. L. 2005, \apj, 629, L105


\bibitem[Caprioli et al.(2009)]{Cea09}
Caprioli, D., Blasi, P., Amato, E., \& Vietri, M. 2009, MNRAS, 395,
895

\bibitem[Caprioli et al.(2010a)]{Cea10con}
Caprioli, D., Amato, E., Blasi, P. 2010, Astropart. Phys., 33, 160

\bibitem[Caprioli et al.(2010b)]{Cea10fre}
Caprioli, D., Amato, E., Blasi, P., 2010, Astropart. Phys., 33, 307

\bibitem[Chevalier \& Liang(1989)]{CL89}
Chevalier, R. A., \& Liang, E. P. \apj, 344, 332

\bibitem[Ellison et al.(2000)]{E00}
Ellison D. C., Berezhko E. G., Baring M. G., 2000, \apj, 540, 292

\bibitem[Fang \& Zhang (2008)]{FZ08}
Fang, J., \& Zhang, L. 2008, \mnras, 384, 1119

\bibitem[Fang et al.(2009)]{Fet09}
Fang, J., Zhang, L., Zhang J. F., Tang, Y. Y. \& Yu, H. 2009,
\mnras, 392, 925

\bibitem[Fang \& Zhang(2010)]{FZ10aa}
Fang, J., \& Zhang, L. 2010, A\&A, in press (arXiv:1003.1656)

\bibitem[Funk et al.(2007)]{Fet07}
Funk, S., et al. 2007, A\&A, 470, 249

\bibitem[Gaensler \& Slane(2006)]{GS06}
Gaensler, B. M., \& Slane, P. O. 2006, ARA\&A, 44, 17

\bibitem[Gelfand et al.(2007)]{Get07}
Gelfand, J. D., Gaensler, B. M., Slane, P. O., Patnaude, D. J.,
Hughes, J. P., \& Camilo, F. 2007, \apj, 663, 468

\bibitem[Gelfand, Slane \& Zhang(2009)]{GSZ09}
Gelfand, J. D., Slane, P. O., \& Zhang, W. 2009, \apj, 703, 2051

\bibitem[Gotthelf \& Halpern(2009)]{GH09}
Gotthelf, E. V., \& Halpern, J. P. 2009, \apj, 700, L158


\bibitem[Helfand et al.(2007)]{Het07}
Helfand, D. J., Gotthelf, E. .V., Halpern, J. P., Camilo, F.,
Semler, D. R., Becker, R. H., \& White, R. L. 2007, \apj, 665, 1297

\bibitem[Kamae et al.(2006)]{Ka06}
Kamae T., Karlsson N., Mizuno T., Abe T., Koi T., 2006, ApJ, 647,
692


\bibitem[Malkov \& Drury(2001)]{MD01}
Malkov M. A., Drury, L. O'C, 2001, Rep. Prog, Phys., 64, 429

\bibitem[Morlino, Amato \& Blasi(2009)]{MAB09}
Morlino, G., Amato, E., \& Blasi, P. 2009, MNRAS, 392, 240

\bibitem[Morlino et al.(2009)]{Mea09}
Morlino, G., Amato, E., Blasi, P., \& Caprioli, D. 2009,
arXiv:0912.2972


\bibitem[Renaud et al.(2009)]{Rea09}
Renaud, M., Marandon, V., Gotthelf, E. V., Rodriguez, J., Terrier,
R., Mattana, F., Lebrun, F., Tomsick, J. A., Manchester, R. N. 2010,
ApJ, 716, 663

\bibitem[Slane(2008)]{S08}
Slane, P. 2008, AIPC, 1085, 120

\bibitem[Slane et al.(2008)]{Set08}
Slane, P., Helfand, D. J., Reynolds, S. P., Gaensler, B.M., Lemiere,
A., \& Wang, Z. 2008, \apj, 676, L33

\bibitem[Slane et al.(2010)]{Slea10}
Slane, P., Castro, D., Funk, S.; Uchiyama, Y., Lemiere, A., Gelfand,
J. D., \& Lemoine-Goumard, M. 2010, (arXiv:1004.2936)

\bibitem[Spitkovsky(2008)]{Sp08}
Spitkovsky, A. 2008, \apj, 682, L5

\bibitem[Truelove \& McKee(1999)]{TM99}
Truelove, J. K., \& McKee, C. F. 1999, ApJS, 120, 299

\bibitem[Ubertini et al.(2005)]{Uet05}                 
Ubertini, P. et al. 2005, \apj, 629, L109

\bibitem[Venter \& de Jager(2006)]{Vd06}
Venter, C., \& de Jager, O. C. 2006, in Proc. 363rdWE-Heraeus
Seminar, Neutron Stars and Pulsars, ed. W. Becker \& H. H. Huang
(MPE Report 291; Garching: MPE), 40


\bibitem[Volpi et al.(2008)]{Vet08}
Volpi, D., Del Zanna, L., Amato, E. \& Bucciantini, N. 2008, A\&A,
485, 337


\bibitem[Zhang, Chen \& Fang(2008)]{ZCF08}
Zhang, L., Chen, S. B., \& Fang, J. 2008, \apj, 676, 1216

\bibitem[Zirakashvili \& Aharonian(2007)]{ZA07}
Zirakashvili, V. N., \& Aharonian, F., 2007, A\&A, 465, 695

\end{thebibliography}
\end{document}